\documentclass[useAMS,usenatbib]{mn2e}

\def\RE{R_{\mathrm{E}}}

\newcommand\ion[2]{#1$\;${\scshape{#2}}}%
\def\Jsdss{{DES J}2146$-$0047}
\def\Jdes{{DES J}0115$-$5244}
\def\J3rd{{DES J}2200$+$0110}
\def\pp{^{\prime\prime}}

\usepackage{lineno}
\usepackage{graphicx}
\usepackage{float}
\usepackage{amssymb}
\usepackage{amsfonts}
\usepackage{amsmath} 
\usepackage{color}
\def\aaemail{\tt aagnello@astro.ucla.edu, tt@astro.ucla.edu}

\def\ucla{Department of Physics and Astronomy, PAB, 430 Portola Plaza, Box 951547, Los Angeles, CA 90095-1547, USA}
\def\ucd{Department of Physics, University of California Davis, 1 Shields Avenue, Davis, CA 95616, USA}
\def\asiaa{Institute of Astronomy and Astrophysics, Academia Sinica, P.O.~Box 23-141, Taipei 10617, Taiwan}
\def\ioa{Institute of Astronomy, Madingley Road, Cambridge CB3 0HA, UK}
\def\kavli{Kavli Institute for Cosmology, University of Cambridge, Madingley Road, Cambridge CB3 0HA, UK}
\def\braz{CAPES Foundation, Ministry of Education of Brazil, Bras{\'i}lia - DF 70040-020, Brazil}
\def\braza{Laborat\'orio Interinstitucional de e-Astronomia - LIneA, Rua Gal. Jos\'e Cristino 77, Rio de Janeiro, RJ - 20921-400, Brazil}
\def\brazb{Observat\'orio Nacional, Rua Gal. Jos\'e Cristino 77, Rio de Janeiro, RJ - 20921-400, Brazil}
\def\brazc{Departamento de F\'{\i}sica Matem\'atica,  Instituto de F\'{\i}sica, Universidade de S\~ao Paulo,  CP 66318, CEP 05314-970, S\~ao Paulo, SP,  Brazil}
\def\brazd{Instituto de F\'\i sica, UFRGS, Caixa Postal 15051, Porto Alegre, RS - 91501-970, Brazil}
\def\ipmu{Kavli IPMU (WPI), UTIAS, The University of Tokyo, Kashiwa, Chiba 277-8583, Japan}
\def\mit{MIT Kavli Institute for Astrophysics and Space Research, 37-664G, 77 Massachusetts Avenue, Cambridge, MA 02139}
\def\fnal{Fermi National Accelerator Laboratory, Batavia, IL 60510}
\def\epfl{Laboratoire d'Astrophysique, Ecole Polytechnique F\'ed\'erale de Lausanne (EPFL), Observatoire de Sauverny, CH-1290 Versoix, Switzerland}
\def\slac{Kavli Institute for Particle Astrophysics and Cosmology, Stanford University, 452 Lomita Mall, Stanford, CA 94035, USA}
\def\illa{Department of Astronomy, University of Illinois, 1002 W. Green Street, Urbana, IL 61801, USA}
\def\illb{National Center for Supercomputing Applications, 1205 West Clark St., Urbana, IL 61801, USA}
\def\espa{Institut de Ci\`encies de l'Espai, IEEC-CSIC, Campus UAB, Carrer de Can Magrans, s/n,  08193 Bellaterra, Barcelona, Spain}
\def\espb{Institut de F\'{\i}sica d'Altes Energies, Universitat Aut\`onoma de Barcelona, E-08193 Bellaterra, Barcelona, Spain}
\def\cata{Instituci\'o Catalana de Recerca i Estudis Avan\c{c}ats, E-08010 Barcelona, Spain}
\def\espc{Centro de Investigaciones Energ\'eticas, Medioambientales y Tecnol\'ogicas (CIEMAT), Madrid, Spain}
\def\exc{Excellence Cluster Universe, Boltzmannstr.\ 2, 85748 Garching, Germany}
\def\lmu{Faculty of Physics, Ludwig-Maximilians University, Scheinerstr. 1, 81679 Munich, Germany}
\def\stern{Universit\"ats-Sternwarte, Fakult\"at f\"ur Physik, Ludwig-Maximilians Universit\"at M\"unchen, Scheinerstr. 1, 81679 M\"unchen, Germany}
\def\penns{Department of Physics and Astronomy, University of Pennsylvania, Philadelphia, PA 19104, USA}
\def\jpl{Jet Propulsion Laboratory, California Institute of Technology, 4800 Oak Grove Dr., Pasadena, CA 91109, USA}
\def\mich{Department of Physics, University of Michigan, Ann Arbor, MI 48109, USA}
\def\micha{Department of Astronomy, University of Michigan, Ann Arbor, MI 48109, USA}
\def\mpe{Max Planck Institute for Extraterrestrial Physics, Giessenbachstrasse, 85748 Garching, Germany}
\def\ohioa{Center for Cosmology and Astro-Particle Physics, The Ohio State University, Columbus, OH 43210, USA}
\def\ohiob{Department of Physics, The Ohio State University, Columbus, OH 43210, USA}
\def\aao{Australian Astronomical Observatory, North Ryde, NSW 2113, Australia}
\def\texas{George P. and Cynthia Woods Mitchell Institute for Fundamental Physics and Astronomy, and Department of Physics and Astronomy, Texas A\&M University, College Station, TX 77843,  USA}
\def\sussex{Department of Physics and Astronomy, Pevensey Building, University of Sussex, Brighton, BN1 9QH, UK}
\def\stanf{Department of Physics, Stanford University, 382 Via Pueblo Mall, Stanford, CA 94305, USA}
\def\ctio{Cerro Tololo Inter-American Observatory, National Optical Astronomy Observatory, Casilla 603, La Serena, Chile}
\def\ucl{Department of Physics \& Astronomy, University College London, Gower Street, London, WC1E 6BT, UK}
\def\cnrs{CNRS, UMR 7095, Institut d'Astrophysique de Paris, F-75014, Paris, France}
\def\sorb{Sorbonne Universit\'es, UPMC Univ Paris 06, UMR 7095, Institut d'Astrophysique de Paris, F-75014, Paris, France}
\def\kicp{Kavli Institute for Particle Astrophysics \& Cosmology, P. O. Box 2450, Stanford University, Stanford, CA 94305, USA}
\def\ports{Institute of Cosmology \& Gravitation, University of Portsmouth, Portsmouth, PO1 3FX, UK}

\title[New lensed quasars in DES]
{Discovery of two gravitationally lensed quasars in the Dark Energy Survey}
\author[Agnello et al.]{
A.~Agnello$^{1,\ast}$,
T.~Treu$^{1,*}$, F.~Ostrovski$^{2,3,4}$, P.L.~Schechter$^{5}$, E.J.~Buckley-Geer$^6$,\and
H.~Lin$^6$,  M.W.~Auger$^2$, F.~Courbin$^7$, C.D.~Fassnacht$^8$, J.~Frieman$^6$, N.~Kuropatkin$^6$, \and
P.J.~Marshall$^9$, R.G.~McMahon$^{2,3}$, G.~Meylan$^7$, A.~More$^{10}$,  S.H.~Suyu$^{11}$, C.E.~Rusu$^8$,\and
D.~Finley$^6$, T.~Abbott$^{12}$,
F.B.~Abdalla$^{13}$,
S.~Allam$^6$,  J.~Annis$^6$, M.~Banerji$^{2,3}$,\and
A.~Benoit-L{\'e}vy$^{13}$, E.~Bertin$^{14,15}$,
D.~Brooks$^{13}$, D.~L.~Burke$^{16,9}$,
A.~Carnero~Rosell$^{17,18}$,\and
M.~Carrasco~Kind$^{19,20}$,
J.~Carretero$^{21,22}$,
C.E.~Cunha$^{16}$, C.B.~D'Andrea$^{23}$,
L.N.~da Costa$^{17,18}$,\and
S.~Desai$^{24,25}$,
H.T.~Diehl$^{6}$, J.P.~Dietrich$^{24,26}$
P.Doel$^{13}$
T.F.~Eifler$^{27,28}$
J.~Estrada$^{6}$,\and
A.~Fausti Neto$^{17}$, B.~Flaugher$^6$, P.~Fosalba$^{21}$,
D.W.~Gerdes$^{29}$,
D.~Gruen$^{30,26}$
G.~Gutierrez$^6$,\and
K.~Honscheid$^{31,32}$,
D.~J.~James$^{12}$, K.~Kuehn$^{33}$,
O.~Lahav$^{13}$, M.~Lima$^{34,17}$,
M.A.G.~Maia$^{17,18}$\and
M.~March$^{27}$, J.L.~Marshall$^{35}$,
P.~Martini$^{31,32}$, P.~Melchior$^{31,32}$,
C.J.~Miller$^{29,36}$,
R.~Miquel$^{37,22}$,\and
R.C.~Nichol$^{23}$, R.~Ogando$^{17,18}$, A.A.~Plazas$^{28}$, K.~Reil$^9$,
A.K.~Romer$^{38}$,
A.~Roodman$^{16,9}$,\and
M.~Sako$^{27}$, E.~Sanchez$^{39}$,
B.~Santiago$^{40,17}$,
V.~Scarpine$^{6}$, M.~Schubnell$^{29}$, I.~Sevilla-Noarbe$^{39,19}$,\and
R.~C.~Smith$^{12}$,
M.~Soares-Santos$^6$, F.~Sobreira$^{6,17}$, E.~Suchyta$^{31,32}$, M.~E.~C.~Swanson$^{20}$,\and
 G.~Tarle$^{29}$, J.~Thaler$^{19}$, D.~Tucker$^{6}$, A.R.~Walker$^{12}$, R.H.~Wechsler$^{41,16,9}$, Y.~Zhang$^{29}$
  \medskip\\
  This paper includes data gathered with the 6.5m Baade Telescopes located at Las Campanas Observatory, Chile.\\
  Affiliations at the end of the paper.\\
  $^*$ Packard Fellow.\\
  $^\ast$\aaemail
}

\begin{document}

\voffset-.6in

\date{Accepted . Received }

\pagerange{\pageref{firstpage}--\pageref{lastpage}} 

\maketitle

\label{firstpage}
\begin{abstract}
We present spectroscopic confirmation of two new lensed quasars
via data obtained at the 6.5m Magellan/Baade Telescope.  The lens
candidates have been selected from the Dark Energy Survey (DES) and WISE
 based on their multi-band photometry and extended morphology in DES images.
%
Images of {\Jdes}~ show two blue point sources at either side of a
red galaxy. Our long-slit data confirm that both point sources are images of the same quasar at $z_{s}=1.64.$
 The Einstein Radius estimated from the DES images is $0.51\pp.$
{\Jsdss}~ is in the area of overlap between DES and the Sloan Digital
Sky Survey (SDSS). Two blue components are visible in the DES and SDSS
images. The SDSS fiber spectrum shows a quasar component at
$z_{s}=2.38$ and absorption compatible with \ion{Mg}{II} and
\ion{Fe}{II} at $z_{l}=0.799$, which we tentatively associate with the 
foreground lens galaxy. The long-slit Magellan spectra show that the
blue components are resolved images of the same quasar. The Einstein
Radius is $0.68\pp$ corresponding to an enclosed mass of $1.6\times10^{11}\,M_{\odot}.$
 Three other candidates were observed and rejected, two being low-redshift pairs of starburst galaxies,
 and one being a quasar behind a blue star.
These first confirmation results provide an important empirical
validation of the data-mining and model-based selection that is being applied to
the entire DES dataset.
\end{abstract}


\section{Introduction}
\begin{figure*}
       \centering
     \includegraphics[width=0.98\textwidth]{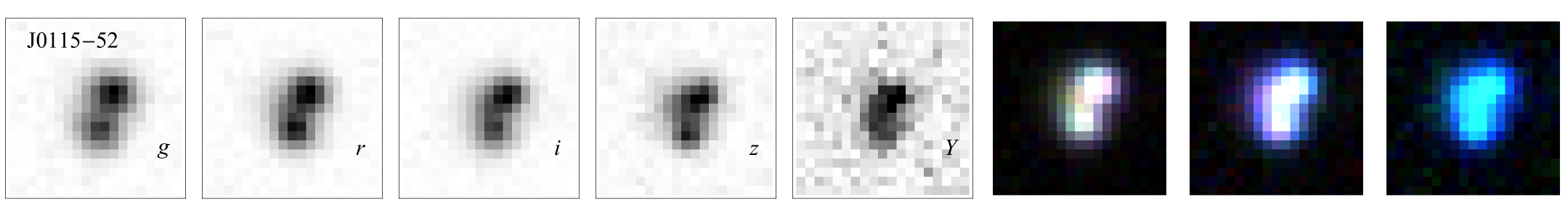}\\
     \includegraphics[width=0.98\textwidth]{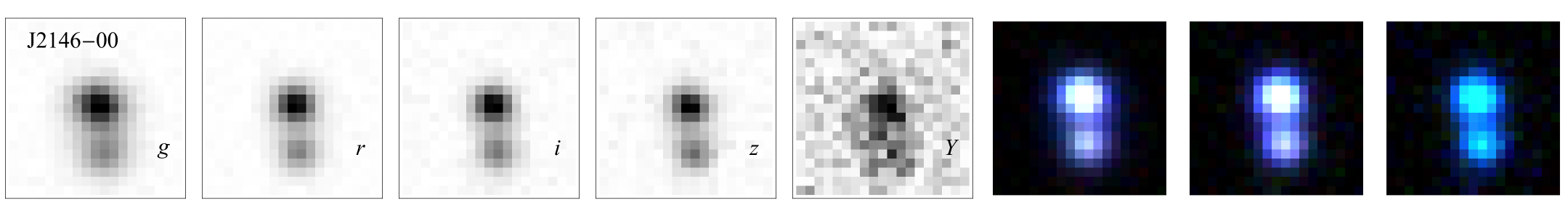}\\

\caption{\small{DES multi-band single-epoch images with best seeing of {\Jdes}~ (top) and {\Jsdss}~ (bottom), in $5^{\prime\prime}\times5^{\prime\prime}$ cutouts. North is up and East is left.
From left to right, DES $g,$ $r,$ $i,$ $z,$ $Y,$ and colour-composites in $gri,$ $riz,$ $izY.$
}}
\label{fig:coadds}
\end{figure*}

Gravitationally lensed quasars provide unique insights into a variety
of fundamental open problems in cosmology and extragalactic
astrophysics \citep[e.g.][]{CSS02}. When a quasar is strongly lensed by a
galaxy, it results in multiple images of the same source, accompanied
by arcs or rings that map the lensed host of the quasar. The light-curves
of different images are offset by a measurable time-delay
\citep[e.g.][]{Sch++97,2013A&A...556A..22T} that depends on the cosmological distances 
to the lens and the source and the gravitational potential of the lens
\citep{Ref64}. This enables one-step measurements of the
expansion history of the Universe and the dark matter halos of the
massive galaxies that act as deflectors \citep[e.g.,][]{Suy++14}. The
microlensing effect on the multiple quasar images, induced by stars in
the deflector, provides a quantitative handle on the stellar content
of the lens galaxies
\citep[e.g.,][]{Schechter:2002p699,2014MNRAS.439.2494O,2014ApJ...793...96S,2015ApJ...799..149J},
and can simultaneously provide constraints on the inner structure of the
lensed quasar, both the accretion disk size and the thermal profile
\citep[e.g.][]{Poi++08,Ang++08,Eig++08,Mot++12}
as well as the geometry of the broad line region (e.g. Sluse et al. 2011,
Guerras et al. 2013, Braibant et al. 2014). Furthermore, milli-lensing
via the so-called flux ratio anomalies provides a unique probe of the
mass function of substructure and thus ultimately of the nature of dark matter
\citep{M+S98,2001ApJ...563....9M,Met02,Dalal:2002p2092,Nie++14}. Finally,
source reconstruction of the lensed quasar and its host give a direct
view of quasar-host coevolution up to $z\sim2$
\citep{Peng:2006p236,2015arXiv150605147R}.
 
Advancement in the field is currently limited by the paucity of known
systems suitable for detailed follow-up and analysis. A large sample
of new systems will be transformative.  To accomplish
this, the STRIDES project\footnote{STRong lensing Insight in the Dark
Energy Survey, PI Treu, full list of Co-PIs and Co-Is at
\texttt{http://strides.physics.ucsb.edu}}, a broad external collaboration of
the Dark Energy Survey\footnote{See \citet{Fla++15} for the technical setup of the Dark Energy Camera.} (DES,
\texttt{http://www.darkenergysurvey.org/index.shtml}), aims at the discovery of the $\sim100$ lensed
 quasars with primary image brighter than i = 21 mag
 predicted by  \citet{O+M10} to lie within the DES footprint.

The identification of such rare systems over 5000 square degrees of
DES imaging data is a classic needle-in-a-haystack problem. The
challenge is possibly greater than that faced in the SDSS dataset
\citep{Ogu++06,Ina++08}, as DES catalogs do not contain u-band data and there is no
built-in spectroscopic dataset to aid in the quasar
selection. Fortunately, the DES image quality (median seeing $\sim 0.9\pp$) is better than that of
SDSS, so one can rely on more accurate morphological information for
candidate selection. New techniques have been developed
in order to address this challenge \citep{agn15,2015ApJ...807..138C}.

Here we report on the first spectroscopic confirmation of lensed quasars from DES. We have obtained spectra of
five of the 68 high-grade, small-separation candidate lensed quasars
in the year-1 DES data release footprint \citep[hereafter Y1A1,][]{DESY1A1},
covering $\approx1200\rm{deg}^{2}$ in the Southern Hemisphere.
 
This \textit{Letter} is organized as
follows. Section~\ref{sect:candsel} briefly illustrates the
candidate-selection process and DES images of the two successful candidates
from the first spectroscopic follow-up. Section~\ref{sect:spectra} shows the long-slit
spectroscopic data obtained for these systems. In
Section~\ref{sect:mods} we present the lensing properties that can be
inferred with current data and conclude in Section~\ref{sect:final}.
Throughout this paper, DES and WISE magnitudes are in the AB and Vega system respectively.
When needed in Sect.~\ref{sect:mods} we adopt a concordance cosmology with $\Omega_{m}=0.3,$
 $\Omega_{\Lambda}=0.7,$ $H_{0}=70\, \rm{km\, s^{-1}Mpc^{-1}}.$
\section{Quasar Lens Candidate Selection}\label{sect:candsel}
\begin{table*} 
\centering
\begin{tabular}{|c|c|c|c|c|c|c|c|c|c|}
\hline
 syst. & $z_s$& $z_l$ & $\Delta$r.a.($\pp$) & $\Delta$dec.($\pp$) & $g$ & $r$ & $i$ & $z$ & $Y$ \\
\hline
\Jdes              & 1.64 & --- & \multicolumn{7}{c}{} \\
   A                 & --- &--- & 0.00 & 0.00 & 20.82$\pm$0.08 & 19.97$\pm$0.10 & 19.91$\pm$0.10 & 20.01$\pm$0.15 & 19.45$\pm$0.14 \\
   B                 & --- &--- & $0.36\pm0.02$ & $0.97\pm0.02$ & 21.12$\pm$0.07 & 20.21$\pm$0.08 & 20.06$\pm$0.10 & 20.21$\pm$0.11 & 19.58$\pm$0.09 \\
   G                 & --- &--- & $0.73\pm0.05$ & $0.22\pm0.05$ & $>$22.66 & 21.35$\pm$0.28 & 20.62$\pm$0.16 & 20.22$\pm$0.13 & 19.59$\pm$0.21 \\
seeing FWHM  & --- & --- & --- & --- & 1.00$\pp$ & 0.98$\pp$ & 0.90$\pp$ & 0.76$\pp$ & 0.75$\pp$ \\
  DES \texttt{model}  & --- &--- & --- & --- & 19.95	&	19.48	&	19.10	&	18.89	&	18.73 \\
\hline
\Jsdss            & 2.38 & 0.799 & \multicolumn{7}{c}{} \\
   A                  & --- &--- & 0.00 & 0.00 & 19.93$\pm$0.13 & 19.85$\pm$0.15 & 19.92$\pm$0.16 & 19.88$\pm$0.16 & 20.23$\pm$0.15 \\
   B                  & --- &--- & -0.17$\pm$0.03 & -1.35$\pm$0.03  & 20.88$\pm$0.14 & 20.50$\pm$0.15 & 20.39$\pm$0.15 & 20.47$\pm$0.15 & 20.71$\pm$0.18 \\
   G                  & --- &--- & 0.17$\pm$0.05 & -0.70$\pm$0.05 & --- & --- & --- & --- & 20.75$\pm$0.30 \\
   G  (CFHT)     & --- &--- & -0.1008$\pm$0.05 & -0.76$\pm$0.05 & --- & --- & 21.04$\pm$0.05 & 20.51$\pm$0.05 & --- \\
seeing FWHM  & --- & --- & --- & --- & 1.12$\pp$ & 0.87$\pp$ & 0.82$\pp$ & 0.77$\pp$ & 1.26$\pp$ \\
 DES \texttt{model}    & --- &--- & --- & --- & 19.52	&	19.67	&	19.48	&	19.16	&	19.05 \\
\hline

\end{tabular}
\caption{Measured parameters of the systems in this paper. The redshift of source $z_s$ and lens $z_l$ are obtained from our long-slit spectra and the SDSS public data. The relative astrometry and magnitudes of brighter (A) and fainter (B) quasar image and lens galaxy (G) are obtained by modelling the single-epoch DES images with best image quality (quoted under `seeing FWHM'). The right ascension (resp.~declination) increases to the E (resp.~to the N).
 Upper limits are given for magnitudes with $\geq0.33$ uncertainty. The galaxy in {\Jsdss}~ is above the detection limit in DES $Y$ band and in CFHT $i,z$ bands.
}
\label{tab:pars}
\end{table*}
We selected small-separation candidates from the DES Y1A1 data
release, using a combination of colour cuts and data mining and
model-based selection. Following \citet{agn15} we adopt a multistep
strategy, whereby {\it targets} are first selected from catalog
data, and {\it candidates} are then selected from the {targets} by
analysis of the actual multiband images. This multistep procedure
allows one to keep the problem computationally fast.

\subsection{Preselection}
Objects are preselected in the DES catalogue based on their `blue' colours and extended morphology.
 The colour selection, satisfying
\begin{eqnarray}
\nonumber  g-r<0.6,\ r-i<0.45,\ i-z<0.55,\\
\nonumber 2.5<i-W1<5.5,\ 0.7<W1-W2<2.0,\\
g-i<1.2(i-W1)-2.8\ ,
\label{eq:cuts}
\end{eqnarray}
is only used to exclude the majority of galaxies and nearby blue stars
from the pool being examined. Of these, we retain those satisfying
\begin{equation}
\mathtt{psf{\_}mag}-\mathtt{model{\_}mag}\geq dmag
\end{equation}
in DES $g,r,i$ bands simultaneously, with $dmag=0.125$ in $g,r$ bands and 0.2 in $i$ band.
 This ensures that the objects are not point-like \citep[cf][who use a similar but converse criterion for point-like objects]{Ree++15}.

Out of $\sim 2\times10^5$ `blue' systems, extended in $i-$band and with acceptable
colours in DES $g,r,i,z$ and WISE $W1,W2$ magnitudes, $\sim 4000$
are brighter than limiting magnitudes $i=21$ or $W1=17$ and  extended also in $g,r$ bands. 

\subsection{Targets}
Of the $4000$ blue extended objects, targets are selected by neural network classifiers
 (ANNs) based on their photometry and multi-band morphology obtained at catalogue-level.
  In particular, DES \texttt{model} $g,r,i,z$ magnitudes and WISE \texttt{w1mpro} ($W1$), \texttt{w2mpro} ($W2$) magnitudes
  are used for the photometry and DES model position angles and axis ratios in $g,r,i,z$ are used for the morphology.
 ANNs produce membership probabilities for each object to belong to different classes, from which the targets are selected
 with cuts in output probabilities. The full procedure is discussed in detail by \citet{agn15}.
ANNs relying just on the photometry select $430$ targets, out of which
 $136$ are also selected by ANNs relying on multi-band morphologies as well.

\subsection{Candidates}
The DES $g,r,i,z,Y$ image cutouts of selected targets have been
modelled as combinations of point sources and galaxies to extract the
SEDs of the extended and point sources. Based on the model results, we
ranked candidates from `0' (not a lens) to `3' (good lens candidate),
based on whether: (i) the point sources have consistent SEDs at least
in $r,i,z$ bands; (ii) their SEDs are compatible with those of quasars;
(iii) the fainter quasar image is also redder in $g-r$; (iv) a possible lens
galaxy is detected in the residuals or in $Y$ band. At this stage, the selection shifts from
 automatic procedures (preselection, ANNs, modelling) to the investigator discretion,
 even though the ranking is based on the four quantitative criteria listed above. The final sample
has 23 grade `3' candidates and 45, 51, 17 with grades `2', `1', `0' respectively.

\subsection{First two confirmed lenses}
Figure~\ref{fig:coadds} shows DES multi-band, single-epoch images in
the best seeing conditions of the two successful candidates for which we
describe spectroscopic follow-up in the next Section. They are {\Jdes}~ at {01:15:57.32 $-$52:44:23.20}
(top) and {\Jsdss}~ at {21:46:46.04 $-$00:47:44.3} (bottom).

 The photometry and relative astrometry of the components in our composite models are summarized in Table~\ref{tab:pars}.
 We have used the single-epoch DES cutouts with best image quality and adopted DES models of the PSF from nearby stars.
 For {\Jsdss}, we also fit Moffat profiles with the same structural parameters to each quasar image and a S{\'e}rsic profile to the lensing galaxy in archival 
 CFHT MegaCam $i,z$-band images (Programme 10BC22, PI: van Waerbeke).
 The resulting photometry is given in Table~\ref{tab:pars} below the DES one.
 Even though the galaxy is below the detection limit in DES single-epoch data except in $Y-$band,
 it is securely detected in CFHT data in $i,z$ bands as well.

Three additional candidates have been observed during the same observing run and ruled out as lensed quasars.
 They are briefly described in Sect.~\ref{sect:falpos} below.

\begin{figure}
       \centering
     \includegraphics[width=0.45\textwidth]{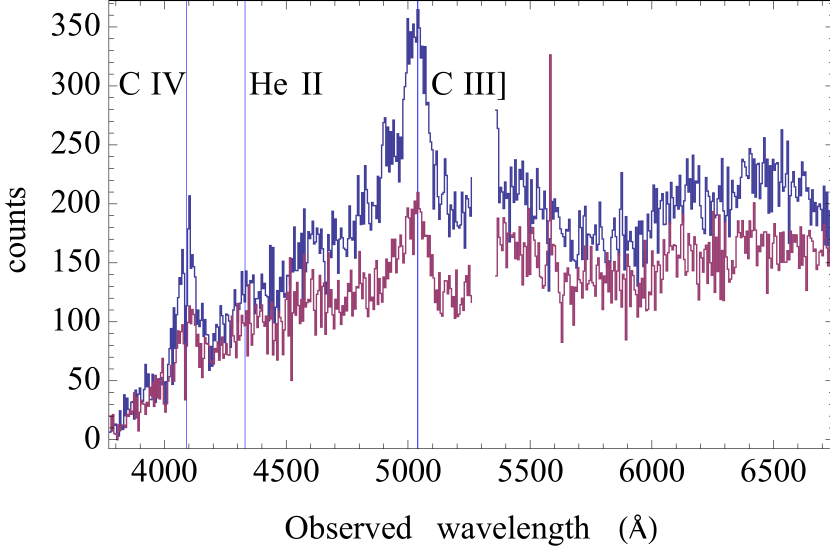}\\
     \includegraphics[width=0.45\textwidth]{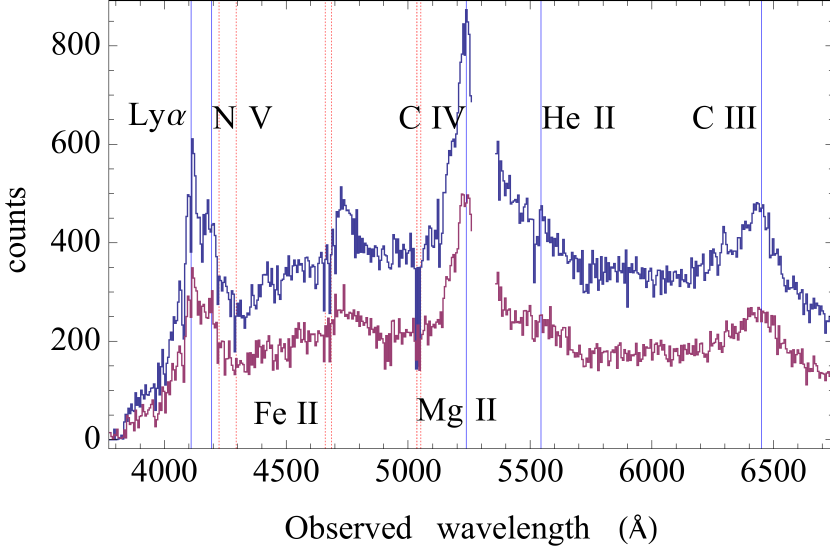}\\

\caption{\small{1D sky-subtracted spectra of {\Jdes}~ (\textit{top}) and {\Jsdss}~ (\textit{bottom}). The red and blue histograms show the spectra of the two components, demonstrating that they are images of the same quasar. Emission and absorption lines are marked by blue (solid) and red (dashed) vertical lines, respectively.
}}
\label{fig:specs}
\end{figure}
\section{Spectroscopic Confirmation}
\label{sect:spectra}
Spectra were obtained with the Inamori Magellan Areal Camera and
Spectrograph \citep[IMACS,][]{dre11} on the Baade 6.5 m telescope on
UT 2015 June 19. The f/4 camera was used with $0.70\pp$ slit mask.
 The data were binned factors of 2 along the slit, giving a scale
of $0.22\pp$/pixel, and 4 in the spectral direction, which combined
with a 300 l/mm grating gave a dispersion of 2.92{\AA}~/pixel.  Two
CCDs covered the spectra over 3770 -- 5260 {\AA}~ and 5350 -- 6880
{\AA}.
  The seeing ranged between $0.8\pp$ and $1.2\pp$, and all our candidates were partially or completely resolved.
  Two spectra of 600s were taken for each object.
 The data were reduced using standard {\sc IRAF }routines.
 One-dimensional (1D) spectra have been extracted by modelling the 2D spectra as superpositions of Gaussian tracks in the spatial dimension, one per component, having peak positions ($p_{1},p_{2}$) that are linear functions of the wavelength
  with the same slope ($\mathrm{d}p/\mathrm{d}\lambda$). Even though the individual tracks are generally well separated,
 this procedure ensures that the resulting 1D spectra are as independent as possible from one another and
 exploit all the information available in the 2D tracks.
 
 Five grade-`3' candidates were visible and observed. The resulting spectra are described below.
 The results on the two confirmed lenses are shown in Fig.~\ref{fig:specs}.

\subsection{\Jdes} 
The 1D spectra of {\Jdes}~ show the same broad emission lines (C IV,
He II, C III$\left.\right]$) at $z_{s}=1.64,$ with a uniform ratio
between the two spectra in the red, as shown in Figure~\ref{fig:specs},
 indicating that the low-order differences between the two spectra are
 due to differential reddening, or perhaps chromatic microlensing.
Together with the presence of a red galaxy in the DES images, the
spectra confirm {\Jdes}~ as a strongly lensed quasar, with image
separation $\approx1.04\pp$.  Unfortunately, the S/N-ratio is too low
to securely detect stellar absorption lines from the deflector.
%
%

\subsection{\Jsdss}
The 1D spectra of {\Jsdss}~ show two components
with the same broad emission lines at $z_{s}=2.38,$ consistent with
the public SDSS fiber spectrum. Prominent Mg II and Fe II absorption lines at
$z_{l}=0.799$ are detected in both spectra (fig.~\ref{fig:specs}). 
 The ratio between the two spectra is constant, which together with the detection of a galaxy
 in $izY$ bands confirms {\Jsdss}~ as a strongly lensed quasar,
 with image separation $\approx1.32\pp.$
 We associate the lens redshift with the galaxy responsible for the absorption lines at $z_{l}.$
 This system has been independently identified as part of the SDSS-III quasar lens sample (More et al., in prep).

\subsection{False Positives}\label{sect:falpos}

 Two candidates, at 20:53:56.5 $-$56:09:36 and 22:17:52.5 $-$53:57:15 respectively,
 are pairs of compact, star forming galaxies. The third rejected candidate, at 22:00:24.11 +01:10:37.56,
 is an alignment of a $z=1.37$ red quasar and a blue star, with the same $r-i$ and $i-z$ colours by coincidence.
 
 In general, pairs of compact star forming galaxies at $z\approx$0.2--0.3 are the main residual contaminant
 from the candidate selection procedure, because their broad-band colours are consistent with those of quasars.
 This includes initial candidates that we rejected based on their available SDSS spectra.
 In WISE magnitudes, they tend to lie at $W2>14$ and $W1-W2<0.8,$ across the limiting locus of \citet{Ass++13},
 a region that however is occupied also by spectroscopically-confirmed quasars and some of the candidates from this search.

\section{Lensing properties}\label{sect:mods}

Even though the current data are not sufficient for a detailed lensing
model, they can be used to derive simple properties of the lens
galaxies. To this aim, we adopt a Singular Isothermal Sphere (SIS)
model for the mass density profile of the deflector, which is commonly
considered the simplest model apt to describe galaxy-scale lenses
\citep{Tre10}.  The SIS projected surface density is
 \begin{equation}
 \Sigma(R)=\frac{1}{2}\Sigma_{cr}(R/\RE)^{-1}
 \end{equation}
 with $\Sigma_{cr}=c^{2}D_{s}/(4\pi G D_{l}D_{ls})$ in terms of
 angular-diameter distances to the source ($D_{s}$), to the lens ($D_{l}$) and
 between lens and source ($D_{ls}$), and the Einstein Radius $\RE$ is
 defined such that it encloses a projected mass
 \begin{equation}
 M_{\rm E}\ =\ \pi\Sigma_{cr}\RE^{2}\ .
 \end{equation}
 The expected line-of-sight velocity dispersion of stars in the lens
 galaxy can be estimated via
\begin{equation}
\sigma^{2}_{\rm{sis}}=\frac{c^{2}\RE D_{s}}{4\pi D_{l}D_{ls}}\ .
\end{equation}
Finally, neglecting microlensing, the total magnification of the point
source can be estimated from the flux ratio $f_{B}/f_{A}$ as
\begin{equation}
\mu_{\rm{tot}}=2\frac{1+f_{B}/f_{A}}{1-f_{B}/f_{A}}\ ,
\end{equation}
where A is the brighter quasar image. Then, from Table~\ref{tab:pars}
and confirmation spectra, we can estimate $\RE,$ $M_{\rm E},$
$\sigma_{\rm{sis}}$ and $\mu_{\rm{tot}}.$ The Einstein radius is
approximated by half the image separation (an exact result for SIS
profiles). For the flux ratio, we use the average between $i$ and $z$ band, for which
the photometric fit is more robust against PSF mismatch and
signal-to-noise ratio, and the effects of dust and microlensing are
minimized. The results are summarized in Table~\ref{tab:lenspars}. For
{\Jdes}, we adopted a lens redshift $z_{l}=0.635,$ which minimizes the
critical surface density $\Sigma_{\rm cr}$ and yields a lower bound on $\sigma_{\rm sis}.$
 We also list the magnification inferred from the ratio of continua in the spectra around 6000~\AA.
 With the positional uncertainties from Table~\ref{tab:pars}, the relative uncertainty on $\RE$ is 1\%.
 With current data, the uncertainties on the magnifications are of the same order as their average value.

\begin{table} 
\centering
\begin{tabular}{|c|c|c|}
\hline
    &  \Jdes   & \Jsdss\\
\hline
$\RE$  & 0.513$\pp$ & 0.680$\pp$ \\
$\mu_{\rm{tot},i}$  & 26.0$\pm$20.0 & 8.4$\pm$3.2 \\
$\mu_{\rm{tot},c}$  & 16.0$\pm$13.0 & 6.6$\pm$1.4 \\
$\sigma_{\rm{sis}}$ (km/s)  & 190 & 215 \\
$M_{\rm E}$ ($M_{\odot}$)    & $0.95\times10^{11}$ & $1.6\times10^{11}$ \\
\hline
\end{tabular}
\caption{Estimated lensing parameters and properties of the lens galaxy:
Einstein radius $\RE$ (in angular units) and total magnification $\mu_{{\rm tot},i}$
 from the fit to the quasar image positions and fluxes; magnification $\mu_{{\rm tot},c}$  on the quasar continua,
  from the 1D spectra; and SIS velocity dipersion $\sigma_{\rm{sis}}$ and enclosed mass $M_{\rm E}$ from the measured redshifts.
  For {\Jdes}, $\sigma_{\rm{sis}}$ and $M_{\rm E}$ are under the hypothesis that the lens is midway between observer and source (i.e. $z_{l}=0.635$).}
\label{tab:lenspars}
\end{table}
\section{Conclusions}
\label{sect:final}
We have presented the very first results of a campaign to extend the
known samples of lensed quasars by exploiting the large footprint of
wide-field photometric surveys. In particular, we have
spectroscopically confirmed two new systems, found in the DES Y1A1
release with data mining techniques. {Two additional candidates were
ruled out by spectroscopic observations as pairs of $z\approx0.2$ compact, narrow-line galaxies.}
Another candidate was ruled out as an alignment of a quasar and a blue star.
  The basic characteristics of the two confirmed systems are as follows.

\textbf{\Jdes}~ 
Consists of two images of a quasar at $z_{s}=1.64$ lensed by a
foreground galaxy visible in the DES images. No redshift is available
for the lens galaxy. The Einstein Radius is estimated to be $0.51\pp$.

\textbf{\Jsdss}~ 
Consists of two images of a quasar at $z_{s}=2.38,$ with prominent Mg
II absorption at $z_{l}=0.799,$ which we tentatively associate with
the lens galaxy redshift. The redshift of the lensing galaxy is not confirmed in
 the spectroscopic follow-up by More et al. (in prep), since detection of the Mg II and Fe II lines
 depends on S/N ratio and in fact is less evident in the worse of our two exposures, where track deconvolution becomes more noisy.
 Thus, deeper spectroscopic data will be needed to measure the redshift of the deflector, possibly based on
 stellar absorption lines. The Einstein Radius is estimated to be $0.68\pp$.

The main class of contaminants (including candidates rejected based on their SDSS spectra)
 consists of groups of compact, star-forming galaxies at $z\sim0.2-0.3,$ because of their broad-band colours
  and compact morphology. With the observed sample, a strict cut in WISE $W1-W2$ vs $W2$ would give a $100\%$
  success rate, which drops at $40\%$ when the cut is relaxed. Still, with just five systems
  currently at hand, we would rather caution against these simple estimates.

Our search has delivered over 100 additional candidates from the Y1A1
data -- and we expect many more from the next seasons of DES data.
 The results of this first follow-up effort are encouraging. However, a systematic
follow-up campaign is needed to confirm large numbers of candidates,
assess the purity of our selection technique, and carry out the many
scientific investigations enabled by lensed quasars.

\section*{Acknowledgments} 
AA, TT, CDF and CER acknowledge support from NSF grants AST-1312329 and
AST-1450141 ``Collaborative Research: Accurate cosmology with strong
gravitational lens time delays''. AA, and TT gratefully acknowledge
support by the Packard Foundation through a Packard Research
Fellowship to TT.  S.H.S. acknowledges support from the Ministry of
Science and Technology in Taiwan via grant
MOST-103-2112-M-001-003-MY3.  FC and GM are supported by the Swiss
National Science Foundation (SNSF).  The work of PJM was supported by
the U.S. Department of Energy under contract number DE-AC02-76SF00515.
We thank Tamara Davis, Cristina Furlanetto, Gary Bernstein and Tom Collett
 for useful comments on earlier versions of this \textit{Letter}. 

This paper has gone through internal review by the DES collaboration.
Funding for the DES Projects has been provided by the DOE and NSF(USA), MISE(Spain), STFC(UK), HEFCE(UK). NCSA(UIUC), KICP(U. Chicago), CCAPP(Ohio State),  MIFPA(Texas A\&M), CNPQ, FAPERJ, FINEP (Brazil), MINECO(Spain), DFG(Germany) and the Collaborating Institutions in the Dark Energy Survey.  The Collaborating Institutions are Argonne Lab, UC Santa Cruz, University of Cambridge, CIEMAT-Madrid, University of Chicago, University College London,  DES-Brazil Consortium, University of Edinburgh, ETH Z{\"u}rich, Fermilab, University of Illinois, ICE (IEEC-CSIC), IFAE Barcelona, Lawrence Berkeley Lab,  LMU M{\"u}nchen and the associated Excellence Cluster Universe, University of Michigan, NOAO, University of Nottingham, Ohio State University, University of  Pennsylvania, University of Portsmouth, SLAC National Lab, Stanford University, University of Sussex, and Texas A\&M University.  The DES Data Management System is supported by the NSF under Grant Number AST-1138766. The DES participants from Spanish institutions are partially  supported by MINECO under grants AYA2012-39559, ESP2013-48274, FPA2013-47986, and Centro de Excelencia Severo Ochoa SEV-2012-0234. Research leading  to these results has received funding from the ERC under the EU's 7$^{\rm th}$ Framework Programme including grants ERC 240672, 291329 and 306478.
\bibliography{references}

\section*{Affiliations}
{\small
  $^1$\ucla\\
  $^2$\ioa\\
  $^3$\kavli\\
  $^4$\braz\\
  $^5$\mit\\
  $^6$\fnal\\
  $^7$\epfl\\
  $^8$\ucd\\
  $^9$\slac\\
  $^{10}$\ipmu\\
  $^{11}$\asiaa\\
  $^{12}$\ctio\\
  $^{13}$\ucl\\
  $^{14}$\cnrs\\
  $^{15}$\sorb\\
  $^{16}$\kicp\\
  $^{17}$\braza\\
  $^{18}$\brazb\\
  $^{19}$\illa\\
  $^{20}$\illb\\
  $^{21}$\espa\\
  $^{22}$\espb\\
  $^{23}$\ports\\
  $^{24}$\exc\\
  $^{25}$\lmu\\
  $^{26}$\mpe\\
  $^{27}$\penns\\
  $^{28}$\jpl\\
  $^{29}$\mich\\
  $^{30}$\stern\\
  $^{31}$\ohioa\\
  $^{32}$\ohiob\\
  $^{33}$\aao\\
  $^{34}$\brazc\\
  $^{35}$\texas\\
  $^{36}$\micha\\
  $^{37}$\cata\\
  $^{38}$\sussex\\
  $^{39}$\espc\\
  $^{40}$\brazd\\
  $^{41}$\stanf\\
  }

\label{lastpage}
\end{document}